\documentclass[nolinenumbers]{aastex631}
\usepackage{amsmath}
\usepackage{textcomp}
\usepackage{booktabs}
\usepackage{threeparttable}
\usepackage{graphicx}
\usepackage{ulem} 
\usepackage{xcolor}
\usepackage{hyperref}

\newcommand{\E}[1]{\times 10^{#1}}
\newcommand{\lt}{\left}       \newcommand{\rt}{\right}

   \newcommand{\gray}{$\gamma$-ray}

\newcommand{\s}{\,{\rm s}}      \newcommand{\ps}{\,{\rm s}^{-1}}
    \newcommand{\kyr}{\,{\rm kyr}}   
\newcommand{\Msun}{M_{\odot}}
\newcommand{\cm}{\,{\rm cm}}    \newcommand{\km}{\,{\rm km}}

\newcommand{\parsec}{\,{\rm pc}}\newcommand{\kpc}{\,{\rm kpc}} 
\newcommand{\erg}{\,{\rm erg}}        \newcommand{\K}{\,{\rm K}}

\newcommand{\G}{\,{\rm G}}      \newcommand{\muG}{\,\mu{\rm G}}
\newcommand{\Jy}{\,{\rm Jy}}      \newcommand{\sr}{\,{\rm sr}}

\begin{document}

\sloppy
\title{Modeling the Saddle-like GeV--TeV Spectrum of HESS J1809--193: 
$\gamma$-Rays Arising from Reverse-Shocked Pulsar Wind Nebula?}

\author[0009-0000-5615-7769]{Jiaxu Sun}
\affiliation{School of Astronomy \& Space Science, Nanjing University, 163 Xianlin Avenue, Nanjing 210023, China}

\author[0000-0002-4753-2798]{Yang Chen}
\affiliation{School of Astronomy \& Space Science, Nanjing University, 163 Xianlin Avenue, Nanjing 210023, China}
\affiliation{Key Laboratory of Modern Astronomy and Astrophysics, Nanjing University, Ministry of Education, Nanjing 210023, China}

\author[0000-0001-8918-5248]{Yiwei Bao}
\affiliation{Tsung-Dao Lee Institute, Shanghai Jiao Tong University, Shanghai 201210, China}
\affiliation{School of Physics and Astronomy, Shanghai Jiao Tong University, Shanghai 200240, China}

\author[0000-0002-9392-547X]{Xiao Zhang}
\affiliation{School of Physics and Technology, Nanjing Normal University, Nanjing, 210023, Jiangsu, People’s Republic of China}

\author[0000-0003-2418-3350]{Xin Zhou}
\affiliation{Purple Mountain Observatory and Key Laboratory of Radio Astronomy, Chinese Academy of Sciences, 10 Yuanhua Road, Nanjing 210023, China}

\correspondingauthor{Yang Chen, Yiwei Bao, Xiao Zhang}
\email{ygchen@nju.edu.cn, byw19952008@163.com, xiaozhang@njnu.edu.cn}


\begin{abstract}
Evolution of pulsar wind nebulae (PWNe) could be expected to leave imprints in \gray{s}.
We suggest that intriguing GeV-TeV spectral energy distribution (SED) of HESS J1809$-$193 and Fermi-LAT source J1810.3$-$1925e is very likely to be the \gray\ signature of PWN J1809$-$193 in light of the scenario that the PWN was struck by the reverse shock of the parent supernova remnant.
Based on evolutionary theory of PWNe, we consider that, when the PWN was disrupted during collision by the reverse shock, some very high-energy electrons escaped impulsively. 
The remaining electrons stayed in the relic PWN, which was displaced from the pulsar.
The very high-energy part of the remaining electrons were depleted by the strong magnetic field that was enhanced by the reverse shock compression in the reverberation stage, leaving the other part of them generating GeV emission. 
The particles injected from the pulsar after the disruption enter the relic PWN through the newly formed tunnel called the cocoon.
The \gray\ emission from the escaped electrons can account for the TeV spectrum of component A of HESS J1809$-$193 or the TeV halo,
while the electrons remaining after disruption can account for the GeV spectrum of J1810.3$-$1925e.
Thus, combination of contributions from these two populations of electrons naturally reproduces the saddle-like SED of HESS J1809$-$193 and J1810.3$-$1925e from 5 GeV to 30 TeV,
together with the spectral hardening around 100 GeV. 
We also show that the post-disruption injection of electrons can explain the spectrum of the relatively faint \gray\ emission of component B of HESS J1809$-$193.

\end{abstract}



\section{Introduction} \label{sec:intro}
Recent years have witnessed the significant advance in the studies of pulsar wind nebulae (PWNe) and supernova remnants (SNRs) by mutliwavelength (up to ultrahigh energy) observations, which contributes to our understanding of the origin of high energy cosmic rays and \gray\ emissions in the Galaxy. 
However, due to irregular spectral characteristics and complicated distribution of objects in the field of view, the nature of many very high energy (VHE) \gray\ sources remains uncertain. 

In this paper, we discuss the intriguing extended VHE source HESS J1809$-$193 with $\sim 0^{\circ}.62$ in semi-major axis and $e = 0.824$ \citep{2023A&A...672A.103H}.
which is situated in a region rich with potential astrophysical counterparts and exhibits distinct spectral properties.
The source was first identified by H.E.S.S.\ during a Galactic Plane Survey \citep{Aharonian2007} and later resolved into two components (A and B) by \citet{2023A&A...672A.103H}. 
Component~A is extended, exhibiting a spectral cut-off at $\sim 13$\,TeV,
and component~B is compact, showing no clear spectral cut-off.
Analysis of Fermi-LAT data confirmed the presence of an extended source, J1810.3$-$1925e, which
appears to be related to component~A of HESS J1809$-$193 in view of its location and morphology 
 \citep{2023A&A...672A.103H}. 
The region contains several pulsars, SNRs, and molecular clouds, which makes it difficult to identify the exact origin of the \gray{s}. 
Three noticeable pulsars, the transient X-ray magnetar XTE J1810$-$197 \citep{Alford2016}, PSR J1811--1925 (with energy loss rate $ 6.4 \times 10^{36} \, \text{erg s}^{-1}\text{, at distance } d \sim 5$\,kpc  \citep{Aharonian2007}), and PSR J1809$-$1917 (with $1.8 \times 10^{36} \, \text{erg s}^{-1}$ and characteristic age $\tau_c$ = 51.4 kyr, at $d \sim 3.3$\,kpc \citep{Aharonian2007}), are located in this region.
Based on the existing observational results,the \gray\ source HESS J1809$-$193 is suggested to be unrelated to magnetar J1810$-$197 \citep{2022ApJ...931...67M} and PSR J1811$-$1925 along with its wind nebula \citep{2023A&A...672A.103H}. 
Thus PSR J1809$-$1917 stands out to be the most plausible candidate for the origin of the $\gamma$-rays.
PSR J1809$-$1917 powers an X-ray PWN (hereafter PWN J1809$-$193) that spans an angular size around $\sim 0^{\circ}.3$ \citep{Kargaltsev2007, Anada2010, 2018ApJ...868..119K, 2020ApJ...901..157K,2020PhRvL.124b1102A}.
The extended  emission of HESS J1809$-$193 also overlaps with two known SNRs, G011.1+00.1 and G011.0$-$00.0,
but there is no clear evidence of association with either G011.0$-$00.0 or G011.0$+$00.1 \citep{2018ApJ...868..119K,2020ApJ...901..157K,2023A&A...672A.103H}.
HAWC has detected VHE $\gamma$-ray emission from HESS J1809$-$193, with energies exceeding 56 TeV and potentially extending beyond 100 TeV \citep{Abeysekara2020, Goodman2022,albert2024tevanalysissourcerich}, establishing this source as one of the most energetic objects in the TeV range. 
Chandra observations revealed a faint diffuse 
X-ray emission that stretches to the south of compact part of PWN harboring PSR J1809$-$1917 and a bright cocoon-like structure along the central axis of the extended part of PWN \citep{Klingler_2020}.
Fermi-LAT Fermi-LAT shows an extended GeV emission overlapping the TeV $\gamma$-ray emission. 
However, the GeV emission spectrum of the associated source J1810.3$-$1925e cannot connect smoothly to the TeV spectrum of component A of HESS J1809$-$193, implying the need for a spectral hardening around 100 GeV \citep{Araya2018, albert2024tevanalysissourcerich}. 
The TeV emission of HESS J1809$-$193 covers a broader area than the GeV emission \citep{2023A&A...672A.103H}.

Many studies have explored both leptonic and hadronic scenarios for HESS J1809$-$193. 
The discovery paper \citep{Aharonian2007} and follow-up works \citep{2008ICRC....2..815K, Renaud_2008} suggested that the PWN around PSR J1809$-$1917 could explain the TeV emission through a leptonic scenario.
In \citet{2023A&A...672A.103H}, component A is considered to be likely caused by inverse Compton (IC) emission from old electrons that form a halo around the PWN; 
and component B could be connected to either the PWN or the SNR and molecular clouds.
However, it later confronts the aforementioned non-smoothed connection of GeV-TeV spectra.
Current leptonic models, including that utilizing HAWC data, could not accommodate the spectrum of Fermi-LAT source J1810.3$-$1925e below $\sim10$\,GeV \citep{2023A&A...672A.103H, albert2024tevanalysissourcerich}.

On the other hand, some studies proposed hadronic interpretations for the  emission from the HESS J1809$-$193 region. Radio observations at 330 MHz and 1456 MHz revealed two SNRs, G011.1+00.1 (10\arcmin  \ in angular diameter) and G011.0$-$00.0 (11\arcmin \ in angular diameter), in the region \citep{Green2004, Brogan2006, Castelletti2016}. 
\citet{Castelletti2016}, \citet{Araya2018}, \citet{Voisin2019}, and \citet{boxi2023hessj1809193gammarayemission}  proposed that cosmic rays accelerated by SNR G011.0$-$00.0 interacting with molecular clouds could be responsible for the emission.
However, a lack of correlation between component A and the gas present in the region disfavors a hadronic interpretation for this component, 
and attributing component B to hadronic explanation would leave the X-ray PWN without a counterpart at TeV energies (component A being attributed to electrons injected long ago only)\citep{2023A&A...672A.103H}.

\gray\  emission from the HESS J1809$-$193 region exhibits spectral hardening in the GeV band and a saddle-like feature in the GeV--TeV range, which is reminiscent of another well-known PWN system, Vela X, which also exhibits a depression around 100 GeV in the GeV--TeV spectral energy distribution (SED) \citep{Tibaldo_2018}. 
The SED of the Vela~X PWN can be interpreted \citep{2019ApJ...877...54B,2019ApJ...881..148B} according to current evolutionary theory of PWNe \citep{Gaensler_2006} and 
hydrodynamic simulation of the Vela X PWN \citep{Slane_2018}.
When the reverse shock from the SNR travels backwards, it encounters the expanding PWN, compressing it. The PWN then experiences 
reverberation, disrupted or even crushed, and very high-energy electrons escape. 
Simultaneously, the compression of the PWN enhances the internal magnetic field,
depleting very high-energy electrons in the relic PWN through radiation losses. 
This process can naturally produce a saddle-featured spectrum with two populations of electrons:
the electrons left in the compressed relic PWN generate GeV emission, while very high-energy electrons escaped from the original PWN at disruption
generate diffusive TeV emission.

In this paper, we suggest that the J1809$-$193 PWN may be another example, next to the Vela~X PWN, producing saddle-like GeV-TeV SED as a result of PWN evolution.
 
\section{\texorpdfstring{MODELLING THE SED OF HESS J1809$-$193}{MODELLING THE SED OF HESS J1809-1917}} \label{sec:model}
\subsection{\texorpdfstring{About structure and evolution of PWN J1809$-$193}{About structure and evolution of PWN J1809-1917}}

\label{sec:HD}
In X-rays (0.5--8\,keV), 
PWN J1809$-$193 is mainly comprised of a compact nebula immediately surrounding the pulsar and an extended nebula in the southwest \citep{Klingler_2020}.
The PWN, $\ga 10'$ in size \citep{Li_2023}, is approximately symmetric about the northeast-southweswest oriented central axis, 
and there is a short (probably by projection) bright bar-like structure, connecting with the compact nebula, 
largely oriented along the axis
 \citep[see Figure~6 in][]{Klingler_2020}.
Such a morphology is consistent with a PWN that SNR reverse shock has swept over and compressed \citep{Gaensler_2006, Slane_2018}.
The bar-like structure is very similar to the ``cocoon'' in the Vela~X PWN \citep{Slane_2018}. 
The extended GeV emission region, which was revealed from the Fermi-LAT observation and is represented by a disk model with a radius of about $\sim 0^{\circ}.3$ \citep{Araya2018} or $8\parsec$ at $d\sim3.3\kpc$,
may correspond to the relic PWN after the passage of reverse shock.
The X-ray emitting part of the PWN is now projected inside this region,
but the tunnel between the pulsar and the relic PWN, represented by the bar-like structure or ``cocoon'', 
needs not perfectly align with the line between the pulsar and the center of the relic PWN, like the case of the PWN in SNR G327.1–1.1 due to addition of the transverse component of the pulsar's proper motion \citep[see Figure 3 in][]{Gaensler_2006,2015ApJ...808..100T}.

According to the hydrodynamic simulation of PWN evolution \citep{Gaensler_2006,2015ApJ...808..100T,Slane_2018}, it can be envisaged that the parent SNR of PWN J1809$-$193 was born in an interstellar environment with a density gradient.
The reverse shock moved inwards from the side with relatively dense ambient medium first collided the PWN, one-sidedly compressing it, and created a trail behind the pulsar.
In the reverberation stage of interaction with reverse shock, the PWN was disrupted, and a small portion of the high-energy particles, which was uncompressed in the trail, impulsively escaped from the PWN into the surrounding region, forming the observed extended TeV halo. 
Meanwhile, most part of high energy particles remained in the relic PWN that was driven to the other side. 
The very high-energy part of the remaining electrons were burnt off by the strong magnetic field that was enhanced by the compression in a short timescale \(\sim 10 (B/{10^2 \muG })^{-2} \left(E/{10^2\, \text{TeV}}\right)^{-1} \, \text{yr}\) (where $E$ denotes the energy of electron) \citep{2011ApJ...743L...7H}.
The other remaining electrons generate the GeV emission.
With the passage of the reverse shock, a compact nebula enveloping the pulsar and a trailing part appeared, 
while a cocoon was formed as a tunnel for subsequent particle injection from the pulsar to the relic PWN.

\begin{figure}[htbp]
    \centering
    \includegraphics[width=0.7\textwidth]{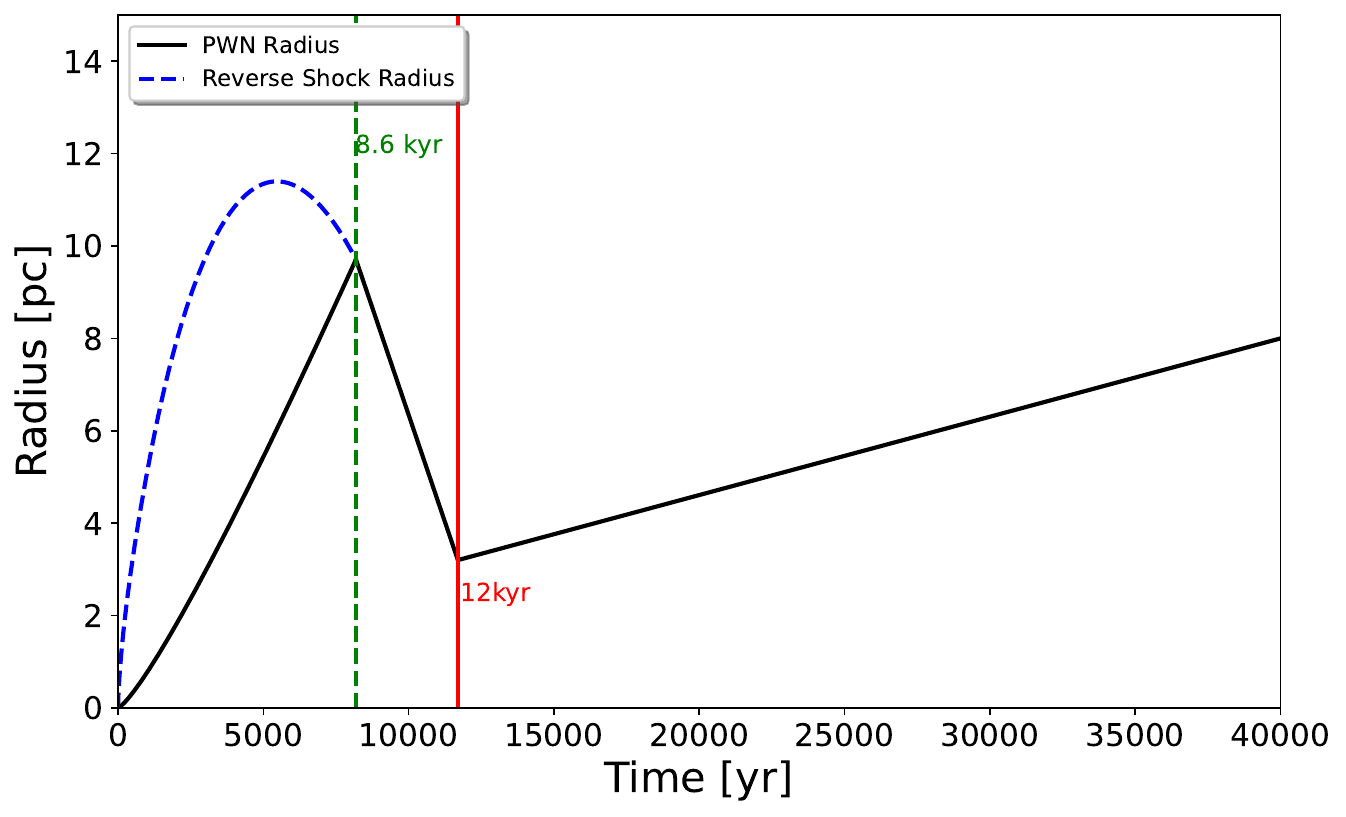}
    \caption{The evolution of the reverse shock radius and the pulsar wind nebula (PWN) radius over time. The green line indicates \( t = 8.6\kyr \), where the interaction begins leading to the compression of the PWN. The red line indicates the characteristic time of disruption \( t_\text{disr} = 12\kyr \).}
    \label{fig:fig.1}
\end{figure}

Figure \ref{fig:fig.1} shows that PWN J1809$-$193  was hit by the reverse shock at $\sim8.6\kyr$ and was compressed to a minimum size at $\sim12\kyr$, 
as calculated according to \citet{1999ApJS..120..299T} 
(also see relevant algorithm described in \citet{2019ApJ...881..148B}),
with the supernova explosion energy $E_{\rm SN}\sim10^{51}\erg$ and 
ejecta mass $M_{\rm ej}\sim5\Msun$ assumed and the ambient gas density $n_\text{ISM}$$\sim0.04\cm^{-3}$ fitted from the evolution of the relic PWN.
After the hitting, the first compression is believed to be the most significant \citep{Bandiera_2023}, particles' escape is assumed to happen around the moment the PWN was compressed to the maximum extent (namely $t_{\rm disr}\sim12\kyr$), 
and the subsequent reverberations may not occur because the PWN lacks the power to re-expand the interface \citep{bandiera2020reverberation}. 
Therefore, this work considers only one complete reverberation cycle.
For the expansion after the compression, we assume that the relic PWN adiabatically expanded at a constant velocity calculated from the balance between PWN and outer pressure following the treatment in \citet{2019ApJ...881..148B}. 
The age of the remnant $T_{\text{age}}\sim40\kyr$ is obtained from the dynamic calculation  so as for the relic PWN to reach a radial extent of {$8\parsec$} presently.

\subsection{Escaped particles and the TeV halo} \label{sec:halo}

Following \citet{2023A&A...672A.103H}, we adopt the single power law distribution, as written below, 
for the escaped very high energy particles to explain the spectrum of the TeV halo:
\begin{equation}
q_\text{inj}(\gamma) =  
    q_\text{A} \delta(t_\text{dif})\left( \frac{\gamma}{10^7} \right)^{-\alpha\!_\text{A}} ,
    \label{eq:inj}
\end{equation}
where $q_\text{A}$ is the injection constant for the electrons with $\gamma = 10^7$, and $\delta(t_\text{dif})$ is delta function.
The transport of TeV \gray\ emitting electrons that are injected at disruption can be described by the following equation:
\begin{equation}
\frac{\partial}{\partial t_{\text{dif}}}f(\gamma, r, t_{\text{dif}}) = \frac{D(\gamma)}{r^2}\frac{\partial}{\partial r}r^2\frac{\partial}{\partial r}f(\gamma, r, t_{\text{dif}}) + \frac{\partial}{\partial \gamma}(P f) + q_{\text{inj}}(\gamma, t_{\text{dif}}),
\label{eq:diff}
\end{equation}
where \( f(\gamma, r, t_{\text{dif}}) \) denotes the electron distribution function, 
\( r \) the radial distance from the center of the TeV halo, 
\(t_\text{dif} = t - t_\text{disr}\) the time after the injection, $P$ the radiation energy loss rate , \(q_{\text{inj}}(\gamma, t)\) the electron injection rate (differential number per unit volume per unit time), and 
\( D(\gamma) = D_0 \left({\gamma}/{\gamma_{40\text{TeV}}}\right)^\delta \)  the energy-dependent diffusion coefficient 
(with $D_0$ the diffusion coefficient normalized at 
Lorentz factor \(\gamma_{40\text{TeV}}\) with 40\,TeV 
and $\delta$ the energy dependence index of diffusion).
The analytical solution to Equation\,\ref{eq:diff} is given by
\begin{equation}
f(\gamma, r, t_\text{dif}) = 
\begin{cases} 
\frac{\gamma_t^2 q_{\text{inj}}(\gamma_t, t_\text{dif})}{\gamma^2 \pi^{3/2} r_{\text{dif}}^3} \exp \left(-\frac{r^2}{r_{\text{dif}}^2}\right) & \gamma \leq \gamma_{\text{max}}, \\
0 & \gamma > \gamma_{\text{max}},
\end{cases}
\end{equation}
where \( r_{\text{dif}}(\gamma, t_\text{dif}) = 2\sqrt{D(\gamma)t_\text{dif} [1 - \left({1 - \gamma/\gamma_{\text{max,halo}}})/({1 - \delta}\right)]} \). 
According to \citet{1995PhRvD..52.3265A}, \( \gamma_{\text{max,halo}} = 1/(p_2 t) \), and \( \gamma_t = \gamma/(1 - p_2 t \gamma) \) represents the initial energy of electrons that have cooled to \( \gamma \) after time \( t \) , 
with $p_2$ being a coefficient for the synchrotron and IC loss rate influenced by the strength of the magnetic field $B_\text{halo}$ (see Eq.(15) therein). 
For \( \gamma < 0.5 \gamma_{\text{max}} \), \( r_{\text{dif}} \approx 2\sqrt{D(\gamma)t_\text{dif}} \).

The column density of electrons with energy $\gamma$ at projection radius $\rho$ is 
\begin{equation}
N_{\text{LoS}}(\gamma, \rho, t_{\text{dif}}) = \int_{-\infty}^{\infty} f(\gamma, r, t_{\text{dif}}) dl = \int_{-\infty}^{\infty} \frac{\gamma_t^2 q_{\text{inj}}(\gamma_t, t_\text{dif})}{\gamma^2 \pi^{3/2} r_{\text{dif}}^3} \exp \left(-\frac{l^2 + \rho^2}{r_{\text{dif}}^2}\right) dl = \frac{\gamma_t^2 q_{\text{inj}}(\gamma_t, t_\text{dif})}{\gamma^2 \pi r_{\text{dif}}^2} \exp \left(-\frac{\rho^2}{r_{\text{dif}}^2}\right), 
\end{equation}
where \( l= \sqrt{r^2-\rho^2} \).
For presenting \gray\ flux $F_\text{halo}$ from the region within the projection radial size $R_{\rm halo}$, we integrate \( F_{\text{LS}} \)  over the region after the injection time to get all of the historical contribution:
\begin{equation}
N_\text{halo}\propto\int_{t_{\text{disr}}}^{T_{\text{age}}} \int_{0}^{R_{\rm halo}} N_{\text{LoS}} 2\pi \rho\,d\rho\,dt_{\text{dif}} = \frac{\gamma_t^2 q_\text{A} \left({\gamma_t}/{10^7}\right)^{-\alpha\!_\text{A}}}{\gamma^2} \left[1 - \exp \left(-\frac{D^2}{r_{\text{dif}}^2(\gamma, T_{\text{age}} - t_{\text{cr}})}\right)\right].
\end{equation}

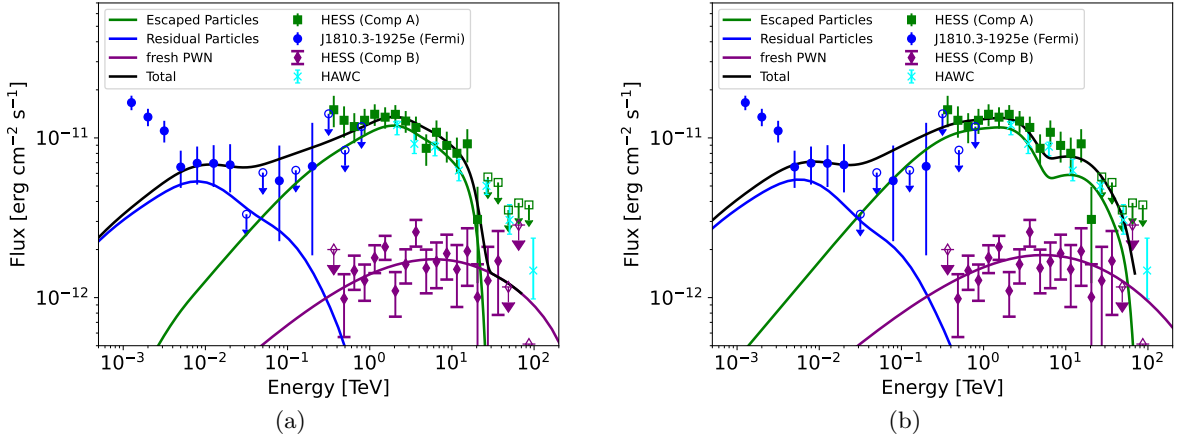
\begin{figure}[htbp]
    \centering
    \begin{minipage}{0.4205\textwidth}
        \centering
        \includegraphics[width=\textwidth]{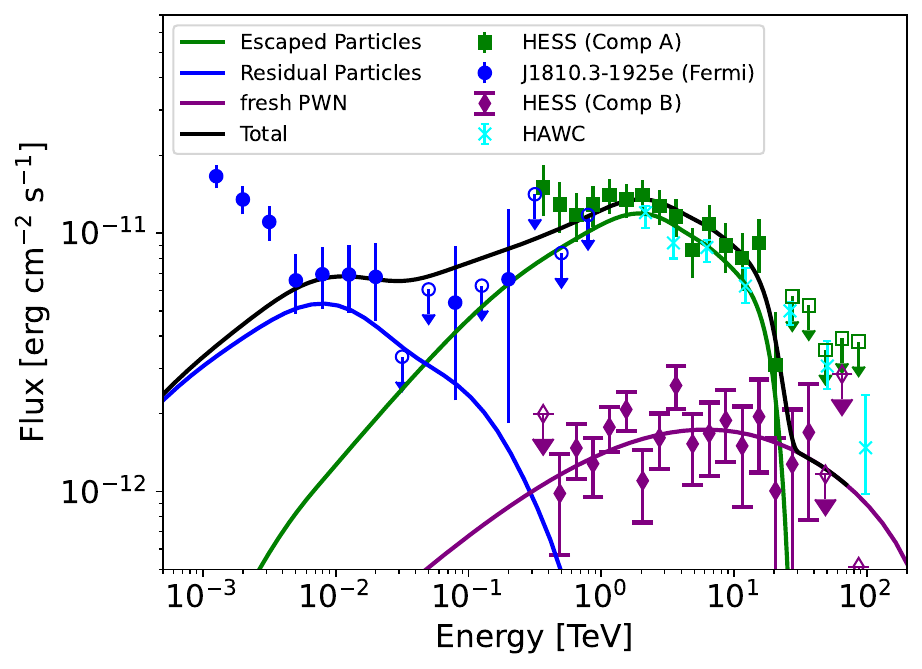}
        \begin{picture}(0,0)
            \put(0,0){\makebox(10,10)[lt]{(a)}}
        \end{picture}
        \label{fig:sed_a}
    \end{minipage}%
    \hspace{0.02\textwidth} 
    \begin{minipage}{0.4205\textwidth}
        \centering
        \includegraphics[width=\textwidth]{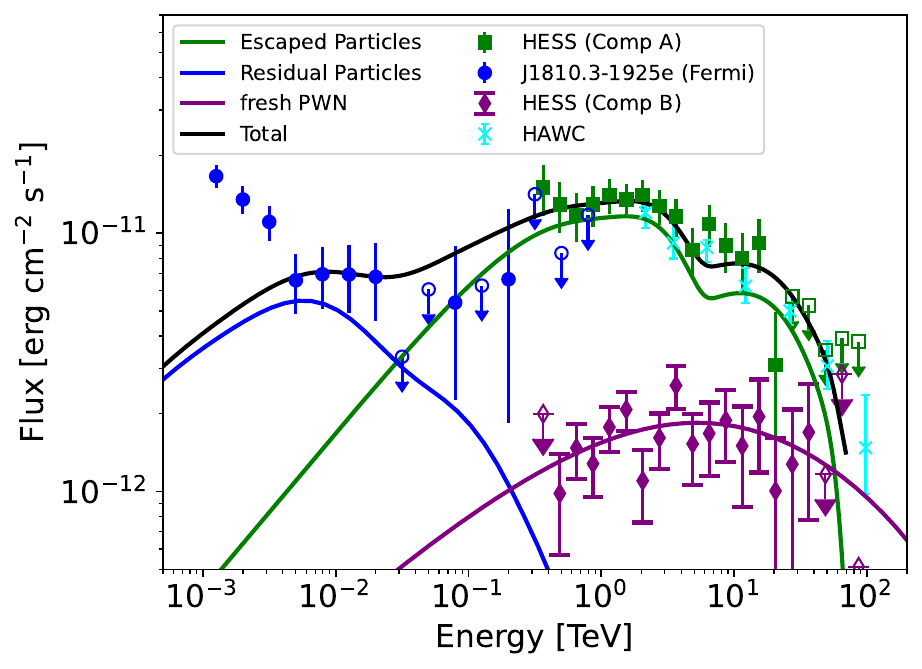}
        \begin{picture}(0,0)
            \put(0,0){\makebox(10,10)[lt]{(b)}}
        \end{picture}
        \label{fig:sed_b}
    \end{minipage}
    \caption{GeV-TeV \gray\ SED of HESS J1809$-$193. 
        Contributions of three components of electrons are shown: (1) electrons escaped at disruption (for HESS component A, in green), remaining electrons after disruption (for Fermi-detected J1810.3$-$1925e arising from relic PWN, in blue), and electrons injected after disruption (for HESS component B, in purple), respectively \citep{Araya2018,2023A&A...672A.103H}.
        The HAWC data points (shown for comparison only) are taken from \citet{Goodman2022}.
        (a) Used parameters are all given in Table\ref{table:table1}.
        (b) Same as in (a), except $\alpha_{A1}=1.75$ and $\alpha_{A2}=1.88$ in the broken power law for injection rate  (Eq.\ref{eq.Halo_broken}), as well as
        $B_\text{Halo} = 2.2 \muG$.
        }
    \label{fig:sed}
\end{figure}

We calculate the diffusion of the very high-energy electrons that escaped ``impulsively" when the reverse-shocked PWN was disrupted and their \gray\ SED.
With the parameter values of $D_0$, $\alpha\!_\text{A}$, $t_\text{disr}$ and $q_\text{A}$ (as listed in Table\,\ref{table:table1}), we calculate the distribution function of the electrons, $f(\gamma, r, t_\text{dif})$. Then we calculate the \gray\ SED of the injected high-energy electrons with parameter values of $B_\text{halo}$, $R_\text{halo}$, $T_\text{age}$, energy densities of intervening FIR, NIR and CMB photons (also see Table\,\ref{table:table1}).
As shown in Figure\,\ref{fig:sed}a, the model calculation can explain the \gray\ fluxes of HESS J1809$-$193 between 0.3\,TeV to 30\,TeV. 
The normalized diffusion coefficient $D_0=1.1\E{28}\cm^2\ps$ is adopted from \citet{2023A&A...672A.103H} and is consistent with the slow diffusion hypothesis in \citet{2017Sci...358..911A}.
Index $\alpha\!_\text{A}$ = 1.8 is the best-fit value here for the TeV emission.
We have also used broken power-law spectrum to fit the SED, but it resulted in little essential difference and improvement below 30 TeV.

Data at the highest energies reported by the HAWC collaboration \citep{albert2024tevanalysissourcerich} will be discussed below (see \S\ref{sec:eHWC}).
\begin{table}[h!]
\centering
\begin{threeparttable}
\caption{Model parameters}
\begin{tabular}{llr}
\toprule
\textbf{Par.} & \textbf{Description} & \textbf{Value} \\
\midrule
\textit{d} (kpc) & distance to the pulsar & 3.3\(^a\) \\
\(L\) (erg s\(^{-1}\)) & pulsar spin-down power & \(1.8 \times 10^{36}\)\(^a\) \\
\(\tau_c\) (kyr) & pulsar characteristic age & 51.4\(^a\) \\
$P$ (ms) & pulsar period & 82.76\(^a\) \\
\(\dot{P}\) (s s\(^{-1}\)) & pulsar period derivative & \(2.55 \times 10^{-14}\)\(^a\) \\
$n$ & pulsar braking index & 3\(^b\) \\
$M_{\rm ej}$ ($M_\odot$) & ejecta mass & 5\(^b\) \\
$E_{\rm SN}$ (erg) & supernova explosion energy & \(10^{51}\)\(^b\)\\
\(\delta\) & energy dependence of the diffusion index & 0.58\(^c\) \\
\(D_0\) (cm\(^2\) s\(^{-1}\)) & diffusion coefficient normalized & \(1.1 \times 10^{28}\)\(^c\) \\
\(T_{\text{NIR}}\) (K) & NIR temperature & 500\,\(^d\) \\
\(u_{\text{NIR}}\) (erg cm\(^{-3}\)) & NIR energy density & \(4 \times 10^{-13}\)\(^d\) \\
\(T_{\text{FIR}}\) (K) & FIR temperature & 31.67\(^d\) \\
\(u_{\text{FIR}}\) (erg cm\(^{-3}\)) & FIR energy density & \(2.05 \times 10^{-12}\)\(^d\) \\
\(T_{\text{CMB}}\) (K) & CMB temperature & 2.72\(^d\) \\
\(u_{\text{CMB}}\) (erg cm\(^{-3}\)) & CMB energy density & \(4.2 \times 10^{-13}\)\(^d\) \\
$B_\text{halo}$ ($\mu$G) & field strength in the TeV halo & 3.5\(^e\) \\
$R_\text{halo}$ (pc) & radius of the TeV halo & 23\(^c\) \\
$q_\text{A}$ ($\text{cm}^{-3}\,\text{s}^{-1}$) & constant of injection into the halo & $1.7 \times 10^{37}$\(^e\) \\
\(\eta\) & magnetic fraction factor before disruption & 0.07\(^e\) \\
$\gamma_{\rm break}$ & break energy of electrons in relic  & $9\times 10^5 $\(^e\) \\
\(\alpha\!_\text{A}\) & power-law index for injection into the halo & 1.8\(^e\) \\
\(\alpha_1\) & index in the broken power law & 1.65\(^e\) \\
\(\alpha_2\) & index in the broken power law & 2.9\(^e\) \\
\(\alpha_\text{B}\) & power-law index for injection after disruption & {2.0\(^e\)} \\
$T_{\rm age}$ (kyr) & age of the pulsar & 40\(^e\) \\
\(n_{\text{ISM}}\) (cm\(^{-3}\)) & ISM number density & 0.04\(^e\) \\
\(\eta_\text{post}\) & magnetic fraction factor after disruption & 0.5\(^e\) \\
$t_{\rm disr}$ (kyr) & time when the PWN was disrupted & 12\(^f\) \\
\bottomrule
\end{tabular}
\begin{tablenotes}
\footnotesize
\item \(^a\) Adopted from \citet{2005AJ....129.1993M}
\item \(^b\) Assumed value
\item \(^c\) Adopted from \citet{2023A&A...672A.103H}
\item \(^d\) Calculated from \citet{2017MNRAS.470.2539P}
\item \(^e\) Fitted value
\item \(^f\) Calculated value
\label{table:table1}
\end{tablenotes}
\end{threeparttable}
\end{table}

\subsection{Remaining particles and the GeV gamma-rays} \label{sec:relic}

While the relic PWN was displaced from the pulsar by reverse shock, the remaining plasma experienced compression
as aforementioned, and the high-energy electrons are postulated burnt off by a strong magnetic field enhanced by compression. Their energies are cut off at the Lorentz factor \(\gamma_{\rm cut}\) due to significant synchrotron losses.  
The distribution function of the remaining electrons in the relic PWN, \(G(\gamma, t)\), is derived by solving the electrons in number conservation equation
\begin{equation}
\frac{\partial G(\gamma, t)}{\partial t} = -\frac{\partial}{\partial \gamma} \left[ \dot{\gamma}(\gamma, t) G(\gamma, t) \right] + Q_{\text{rem}}(\gamma, t) ,
\label{eq:conserve}
\end{equation}
and the injection rate (differential electron number per unit time) of the plasma that were emanated from the pulsar prior to disruption and remained later in the relic PWN is assumed as
\begin{equation}
Q_{\text{rem}}(\gamma, t) = \begin{cases}
    Q_\text{pairs}(\gamma, t)(\gamma/\gamma_\text{break})^{-\alpha_1}  ,&\gamma <\gamma_{\text{break}}  \\
    Q_\text{pairs}(\gamma, t)(\gamma/\gamma_\text{break})^{-\alpha_2}, &\gamma>\gamma_{\text{break}}
\end{cases} ,
\end{equation}where $Q_\text{pairs}$ is a normalization constant and $\alpha_1$ and $\alpha_2$ are the spectral indices of the injected electrons. 
The rate of total electron energy injection into the PWN prior to disruption is expressed as
\begin{equation}
{(1-\eta){L}(t)} = \frac{(1 - \eta) {L}_0}{(1 + \frac{t}{\tau_0})^{(n+1)/(n-1)}} = \int_1^{\gamma_{\text{max,relic}}} Q_{\text{rem}}(\gamma, t)  \gamma {{m_e c^2}}\,d\gamma,\,t \leq t_\text{disr},
\label{eq:inj_relic}
\end{equation}
where \({L}_0\) is the initial spin-down luminosity, 
\(\eta\) the fraction of the spin-down energy deposited into magnetic field, 
$n$ is the braking index, $\tau_0 = 2\tau_c/(n-1)-T_\text{age}$ is the initial spin down age of the pulsar. 
The electron's energy loss rate \(\dot{\gamma}\) in Equation\,\ref{eq:conserve} is determined by synchrotron radiation, IC scattering (with CMB, FIR, and NIR photons), bremsstrahlung, and adiabatic losses.
Integration limit $\gamma_\text{max,relic}$ in Equation\,\ref{eq:inj_relic} is treated similarly to $\gamma_{\rm max,halo}$. 

The magnetic field strength in the relic PWN is governed by the magnetic energy injected and its expansion \citep{2010ApJ...715.1248T}: 
\begin{equation}
\frac{dW_B(t)}{dt} = \eta {L}(t) - \frac{W_B(t)}{R_\text{relic}(t)} \frac{dR_\text{relic}(t)}{dt},\, t \leq t_\text{disr} ,
\label{eqn:field}
\end{equation}
where \(W_B(t)\) is the magnetic energy within the relic PWN, and $R_\text{relic}(t)$ is its radius (which is now 8\,pc, see \S\ref{sec:HD}).

We first calculate the electrons injection rate $Q_\text{rem}$ from Equation  \ref{eq:inj_relic}.
Combining the radius evolution of the relic PWN that is obtained above (see \S\ref{sec:HD} and Figure\,\ref{fig:fig.1}), we get the magnetic field evolution from Equation \ref{eqn:field}, which is used in calculating the energy loss rate $\dot{\gamma}$.
Then we calculate the distribution function $G(\gamma,t)$ of the remaining  particles from Equation \ref{eq:conserve}. 
Finally we obtain the \gray\ SED of the remaining particles from $G(\gamma,t)$, 
incorporating energy densities of intervening FIR, NIR and CMB photons (also see Table\,\ref{table:table1}).
These calculations have used parameters $L$, $\tau_c$, $T_\text{age}$, $P$, $\dot{P}$, $n$, $\eta$, $\alpha_1$, $\alpha_2$, along with energy densities of intervening FIR, NIR and CMB photons (as also listed in Table\,\ref{table:table1}).
As seen in Figure \ref{fig:sed},
the \gray\ emission arising from the remaining particles in the relic PWN is peaked at around 10\,GeV.
Combination of contributions from the relic PWN and the TeV halo can well explain the saddle-like SED of HESS J1809$-$193 from 5 GeV to 30 TeV,
as well as the spectral hardening around 100 GeV. 
We note that, the index $\alpha_1 =1.65$ below the break energy is similar to the index ($\alpha_{\rm A}=1.8$) of the electrons that were escaped into the halo.

In Equation \ref{eq:inj_relic}, the high-energy particles that escaped have been ignored. 
Actually, the total energy of those particles, takes up only $\sim6\%$ of the total energy of the particles injected prior to the disruption in this model calculation.

Regarding the emission below 5 GeV, in our model, the electron energy required to reproduce the observed flux
through the IC process would exceed the pulsar's injection energy, which is evidently unphysical. 
Several authors have proposed that it could originate from non-thermal bremsstrahlung interactions \citep{boxi2023hessj1809193gammarayemission,albert2024tevanalysissourcerich}, which agrees with our findings.




\subsection{Post-disruption injection and HESS ``component B''}
\label{sec:post}
After the PWN was disrupted by the reverse shock, the pulsar continues to inject high energy electrons, which then flow to the relic PWN through the newly formed cocoon.
Hereby, we show that the \gray\ emission from the post-disruption electrons can be responsible for the compact TeV emission ``component B".
We assume that the injection rate of the relativistic particles after disruption, $Q_{\text{post}}(\gamma, t)$, obeys a single power law:
\begin{equation}
Q_{\text{post}}(\gamma, t) =  
Q_\text{B}(t) \gamma^{-\alpha_\text{B}}\hspace{5mm} (\gamma < \gamma_\text{post,max},  t_\text{disr} <t \leq T_\text{age}),
\label{10}
\end{equation}
where $Q_\text{B}$ is the normalization coefficient.
The upper cutoff $\gamma_\text{post,max}$ is obtained so as to confine the accelerated electrons within the PWN \citep{Venter:2006ex}:
\begin{equation}
    \gamma_{\text{post,max}} \approx \frac{e}{2m_e c^2} \sqrt{\frac{\sigma L(t)}{(1+\sigma) c}},
    \label{11}
\end{equation}
where the magnetization parameter $\sigma$ is the ratio of the electromagnetic energy flux to the lepton energy flux and will be approximated here as $ \eta_\text{post}/(1-\eta_\text{post})$, with $\eta_\text{post}$ the fraction of the spin-down energy deposited into magnetic field after disruption.
The injection rate can be related to the spin-down power $L(t)$ at given time $t$ by $(1-\eta_\text{post}) L(t) = \int Q_\text{post}(\gamma, t) \gamma m_e c^2 \, d\gamma$.
Thus the normalization parameter $Q_\text{B}(t)$ can be derived as \citep{2010ApJ...715.1248T}
\begin{equation}
Q_\text{B}(t) = \frac{(1-\eta_\text{post})L_0}{m_e c^2} \left(1 + \frac{t}{\tau_c}\right)^{-2} \lt( \frac{\gamma_{\text{max}}^{2-\alpha_\text{B}} - 1}{2 - \alpha_\text{B}} \rt)^{-1}.
\label{12}
\end{equation}
On the assumption of magnetic field energy conservation, 
the time-varying field strength of the nebula is given by
\citep[see][] {2010ApJ...715.1248T}
\begin{equation}
B(t) = \sqrt{\frac{6\eta_\text{post} L_0 t_0}{R_{\text{PWN}}^3 (t + t_0)}},
\label{eq:B_post}
\end{equation}
where $R_\text{PWN}(t)$ is the mean radius of the PWN.
The volume-integrated particle number $N(\gamma,t)$ as a function of energy and time is described by the continuity equation in the energy space:
\begin{equation}
\frac{\partial N(\gamma, t)}{\partial t}  = -\frac{\partial}{\partial \gamma} \left[ \dot{\gamma}(\gamma, t) N(\gamma, t)  \right] + Q_{\text{post}}(\gamma, t).
\label{15}
\end{equation}
In the above equation, particle escape in this stage is ignored because of longer escape time scale than that of this pulsar system. 

With a similar algorithm to that used by \citet{2010MNRAS.408L..80L}, we have numerically calculated the SED of post-disruption PWN.
Combining equations \ref{10}--\ref{12} and parameters of the PWN, 
we calculate the injection rate $Q_\text{rem}(\gamma,t)$. 
In the calculation, we approximate $R_{\rm PWN}(t)$ as the radius of the re-expanding relic PWN $R_{\rm relic}(t)$ (see Figure\,\ref{fig:fig.1}).
Next, we calculate the magnetic field $B(t)$ via Equation \ref{eq:B_post}.
Then, combined with the parameters of PWN, such as $L$, $\tau_c$, $\alpha_\text{B}$, $\eta_\text{post}$, and energy densities (also see Table \ref{table:table1}), we numerically solved for $N(\gamma,t)$ and calculate the SED of PWN.
As shown in Figure\,\ref{fig:sed}a, the SED of HESS component B can be well fitted.
The injection index $\alpha_{\rm B} =  2$ is slightly larger than $\alpha_{\rm A} = 1.8$. 

On the other hand, hydrodynamic simulations of interaction between reverse shocks and PWNe show that the particles injected from pulsar into the relic PWN (via cocoon) after reverse-shock disruption will form a small central region of relatively high density, close to the injection site. 
This may imply that the emission of the electrons from the post-disruption injection will actually be brightened towards the central region of the relic PWN, appearing as a compact source, as described for HESS component~B.


\section{DISCUSSION} \label{sec:cite}
\subsection{Radio traces of the parent SNR and the relic PWN?}\label{subsec:origin}
Since PWN J1809$-$193 extends significantly beyond the two known nearby SNRs (G011.0$-$00.0 and G011.0$+$00.1), and no SNR spatially coincident with HESS J1809$-$193 has been observed yet, we need to check whether the discussed parent SNR of the PWN is elusive from radio detection due to its expansion in a low-density 
(of order $\sim 10^{-2}\cm^{-3}$) region.
With the Sedov evolutionary law \citep{1958RvMP...30.1077S}, we obtain the radius of the SNR at $T_{\rm age}$ (40\,kyr) as
$R_\text{SNR} = (2.026E_\text{SN}/1.4n_{\rm ISM}m_\text{H})^{1/5}T_{\rm age}^{2/5}\sim$\,40\,pc ($\sim 0^{\circ}.7$ in angular size), 
along with the expansion velocity 
$0.4R_{\rm SNR}/T_{\rm age} \sim 390\km\ps$.

\begin{figure}[htbp]
    \centering
    \begin{minipage}{0.4\textwidth}
        \centering
        \includegraphics[width=\textwidth]{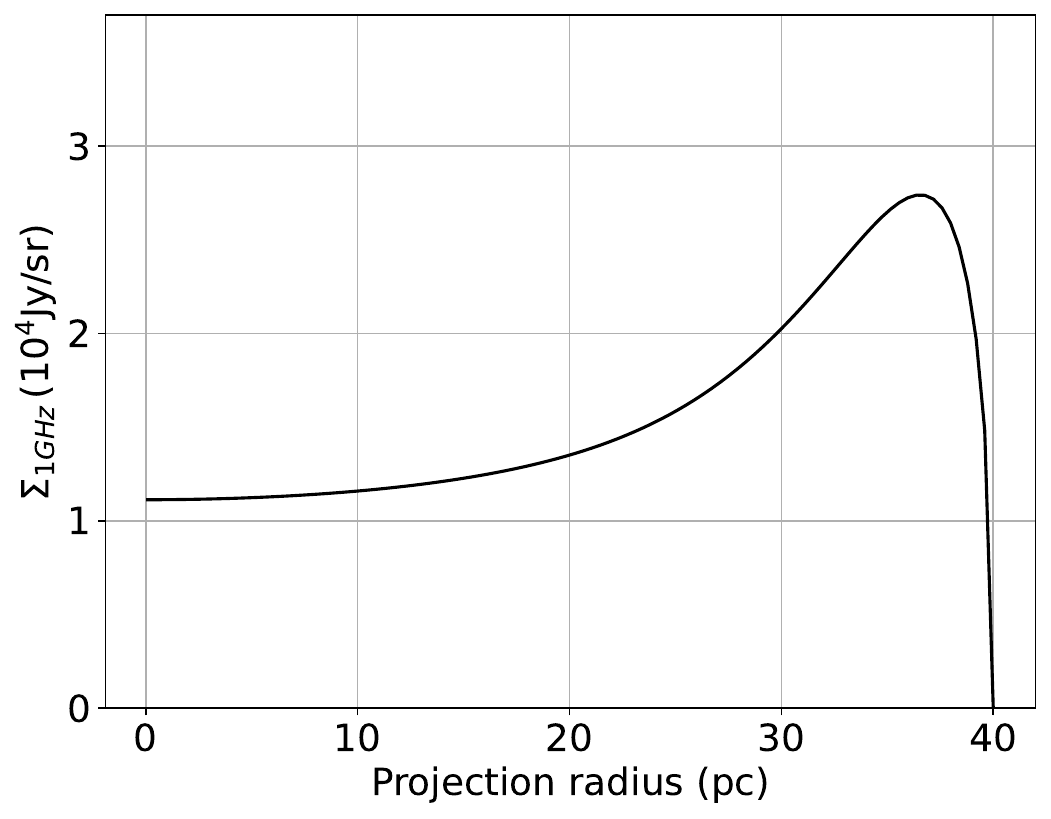}
        \caption{Model-predicted radio surface brightness of the parent SNR}
        \label{fig:radio1}
    \end{minipage}%
    \hspace{0.02\textwidth} 
    \begin{minipage}{0.4205\textwidth}
        \centering
        \includegraphics[width=\textwidth]{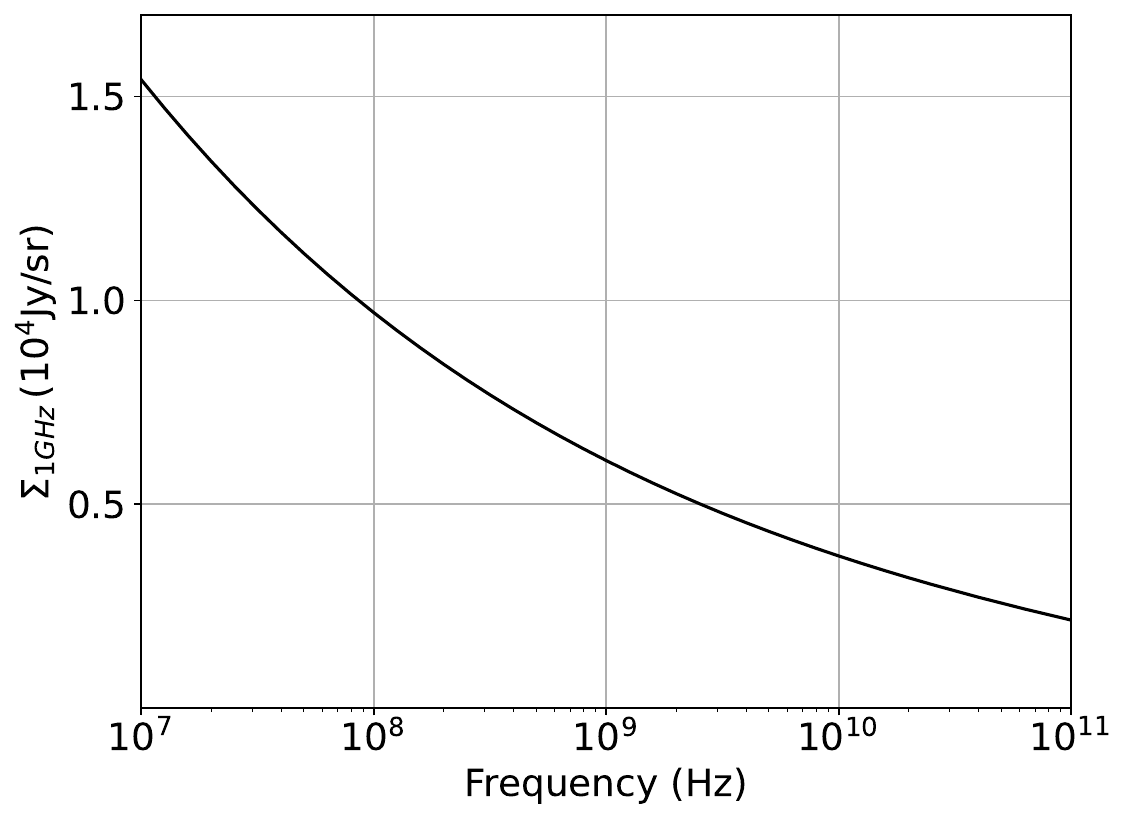}
        \caption{Model-predicted radio surface brightness of the relic PWN}
        \label{fig:radio2}
    \end{minipage}
\end{figure}

We approximate the density distribution with radius for the gas inside the parent SNR as a power law with index 9.
With the synchrotron emission coefficient given in \citet{1970RvMP...42..237B},
we calculate the radio surface brightness distribution $\Sigma$ at 1\,GHz with the projection radius as plotted in Figure\,\ref{fig:radio1},
wherein a total electron energy \(10^{48} \, \text{erg}\), distance $d\sim3.3\kpc$, and a magnetic field strength $10\muG$ are adopted.
The brightness is within a range of $\sim 1$ -- $3\E{4}\Jy\sr^{-1}$.
We alternatively use the \textsc{Naima} package \citep{naima} to estimate the average radio surface brightness of the SNR at 1\,GHz, 
which is $\sim2\E{4}\Jy\sr^{-1}$.

With the scenario of reverse-shocked PWN to account for the \gray\ emission of HESS J1809$-$193, we also need to check whether there may be any visible trace of the relic PWN in radio observation.
Based on the above results regarding the calculation of the electron energy distribution (including the remaining electrons and the post disruption injection) and the relic PWN radius (see \S\ref{sec:HD} and \S\ref{sec:relic}), we can calculate its mean radio surface brightness profile, which is shown in Figure \ref{fig:radio2} to be of order $10^{4}\Jy\sr^{-1}$.

We have estimated the level of radio continuum emission in the background around HESS J1809$-$193, based on the 170 to 231 MHz radio-continuum emission data from the Galactic and Extragalactic All-sky Murchison Widefield Array 
survey.
Four circular regions with faint emission, containing as little enhancement as possible, are selected to estimate the bottom level of background radio continuum emission. The total area of the background regions is comparable to that of the putative SNR.  
The background regions are close to, but outside of, the putative SNR, of which the central Galactic coordinates ($l$ and $b$) and angular radius are ($10^\circ.66$, $0^\circ.97$, $0^\circ.34$), ($10^\circ.85$, $-1^\circ.48$, $0^\circ.34$); ($11^\circ.49$, $1^\circ.00$, $0^\circ.34$), and ($12^\circ.08$, $-0^\circ.96$, $0^\circ.34$), respectively.
The surface brightness of the background is estimated as $\sim5.0\E{5}\Jy\sr^{-1}$. The corresponding extrapolated 1 GHz surface brightness is estimated as $\sim2.2\E{5}\Jy\sr^{-1}$, using a spectral index of $-0.5$.
The radio surface brightness of both of the parent SNR and the relic PWN are lower than the radio background brightness (see Figures 3 and 4).
The variation of the background radio continuum emission is estimated as $\sim2.7\E{5}\Jy\sr^{-1}$, and the extrapolated variation at 1 GHz becomes $\sim1.2\E{5}\Jy\sr^{-1}$.
By comparison, the radio continuum emissions from both the parent SNR and the relic PWN are actually submerged in the background.


\subsection{\texorpdfstring{HAWC detection above 100\,TeV}{HAWC detection above 100 TeV}} \label{sec:eHWC}

Recently, HAWC reported detection of \gray\ flux above 100 TeV with no clear cutoff in the HESS J1809$-$193 region \citep{albert2024tevanalysissourcerich}. 
Such a flux is not explained in the above modeling. 
In the framework of the modeling,
this may indicate that there is also very high-energy particle injection in other ways.
For example, X-ray filaments are observed extending from compact parts of PWNe, such as Guitar Nebula \citep[e.g.][]{2022ApJ...939...70D}, Lighthouse PWN \citep[e.g.][]{Klingler_2023}, etc.
They are interpreted as beams of charge-separated very high-energy electrons or positrons \citep{2024A&A...684L...1O}.
In this scenario,
non-resonant instability 
allows high-energy particles to propagate far from the pulsar along narrow channels in the ISM, forming elongated X-ray filaments
and producing TeV halos that could reach up to 100 TeV or even higher \citep{2023A&A...670A...8M}.
In PWN J1809$-$193 \citep{2020ApJ...901..157K}, a misaligned outflow structure of $\sim7'$ in length was observed 
extending roughly perpendicular to the 
PWN's central axis 
outside the compact nebula. 
It appears similar to the X-ray filaments observed in Guitar Nebula and Lighthouse PWN.
Hence, this outflow structure could provide a potentially feasible explanation for additional high-energy electrons or positrons' escape and generating a \gray\ flux exceeding 100 TeV.

However, the trend seen in HAWC data are similar to the flux level and slope of the TeV emission from component A.
Therefore, it is also possible that the emission fluxes below and above 100\,TeV have a common origin.
In this case, we find that injection rate with a broken power law (instead of a single power law like Eq.\ref{eq:inj}) could be capable of fitting the fluxes.
Using injection rate given by
\begin{equation}
q_\text{inj}(\gamma) = \begin{cases}
    q_\text{A} \delta(t_\text{dif})\left( \frac{\gamma}{10^7} \right)^{-\alpha\!_\text{A1}}  ,&\gamma <\gamma_{\text{break,Halo}}  \\
    q_\text{A} \delta(t_\text{dif})\left( \frac{\gamma}{10^7} \right)^{-\alpha\!_\text{A2}}, &\gamma>\gamma_{\text{break,Halo}}
\end{cases} ,
\label{eq.Halo_broken}
\end{equation}
the SED from GeV to above 100\,TeV is fitted with parameters $\alpha_{A1} = 1.75$, $\alpha_{A2} = 1.88$ and $\gamma_\text{break,Halo} = 1.9\times 10^7$, as well as
magnetic field strength $2.2\muG$ (see Figure\,\ref{fig:sed}b).
Though, this leads to a deviation of the flux in the energy interval 30--200\,GeV. 
Also, the fitted field strength is even somewhat weaker than the interstellar average ($3\muG$), and the break energy $\gamma_\text{break,Halo}$ is an order of magnitude larger than that typically seen in PWNe ($\sim10^6$ \citep{10.1111/j.1365-2966.2010.17449.x}).


\subsection{Comparison with previous explanations}
While pointing out the difficulties with hadronic and lepto-hadronic hybrid models (also see mentioning in \S \ref{sec:intro}), \citet{2023A&A...672A.103H} propose a purely leptonic model to explain the emissions of HESS J1809$-$193.
In the model, HESS component A is ascribed to the electrons injected over the lifetime of the system, 
and component B to `medium-age’ electrons that have been injected within the last $\sim5\kyr$ (without explaining physical origin).
The model requires an additional IC component, emitted by electrons even older than the lifetime to account for the observed $\sim 10$ GeV $\gamma$-ray flux from J1810.3$-$1925e.
In our model with the scenario of reverse-shocked PWN, these three \gray\ components can be explained with the leptonic emissions of three populations of electrons (\S\ref{sec:halo}, \S\ref{sec:relic}, and \S\ref{sec:post}) related to the PWN evolution.
We note that, with different origin mechanism, our calculation of post-disruption electron population described in \S\ref{sec:post} is somewhat similar to the calculation for component B (including power-law index and magnetic energy fraction) in \citet{2023A&A...672A.103H}. 


The lepto-hadronic hybrid explanation proposed by \citet{boxi2023hessj1809193gammarayemission} and \citet{albert2024tevanalysissourcerich} ascribe the TeV halo (HESS component A) to hadronic interaction of SNRs G11.0$-$0.0 and G11.0$+$0.1 with molecular clouds,
and HESS component B to IC scattering of the CMB photons by the PWN electrons.
This model can explain the measured HAWC flux points up to 200 TeV, but seriously overestimates the flux between 5 -- 200\,GeV, of Fermi-LAT detected J1810.3$-$1925e \citep{2023A&A...672A.103H}.
\citet{2024A&A...690A.116M} suggest that component A can be attributed to escape of particles out of
the nebula into the parent remnant and subsequently to the surrounding ISM, and component B can be associated with PWN or hadronic cosmic rays interacting with nearby molecular clouds, however their model needs another component to explain the emission below 10 GeV. 
By comparison to these explanations, our model explanation does not invoke hadronic interaction but only considers the effects of the PWN evolution in the parent SNR.
We do not model the flux beyond 100\,TeV but imply that it is caused by additional injection from the PWN, 
which should be tested with further observations.
\section{Conclusion} \label{sec:summary}
For GeV-TeV source HESS J1809$-$193 together with J1810.3$-$1925e, we suggest that it is very likely to be PWN J1809$-$1917 observed in \gray\ emission.
Based on evolutionary theory of PWNe, we consider that the PWN was collided by the reverse shock moving inwards from the side with relatively dense ambient medium, roughly along the line of sight.
Some very high-energy electrons escaped impulsively when the SNR was disrupted, and the other electrons remained in the relic PWN that was driven to the other side. 
The very high-energy part of the remaining electrons were burnt off by the strong magnetic field that was caused by the reverse shock compression in the reverberation stage, leaving the other part of them generating GeV emission.
The particles injected from the pulsar after the disruption enter the relic PWN through the newly formed tunnel (called the cocoon).
The \gray\ emission arising from the escaped part of electrons can account for the TeV SED of the HESS component A (namely, the TeV halo),
and the electrons remaining after disruption account for the GeV SED emission of J1810.3$-$1925e.
Summation of contributions from these two populations of electrons reproduces the saddle-like SED of HESS J1809$-$193 from 5 GeV to 30 TeV,
as well as the spectral hardening around 100 GeV.
The post-disruption injection of electrons is shown to be responsible for the relatively faint \gray\ emission of HESS component B.
We also show that the radio emission from the PWN's parent SNR, which is assumed to expand in a low-density ISM, and that from the relic PWN are both submerged in the radio background and are thus undetectable.

The emission flux above 100\,TeV newly detected by HAWC from component A could be explained if either a very low field strength (smaller than the interstellar average) and an extraordinarily high break energy of the escaped electrons are invoked, or additional very-high energy injection is present, such as beam-like electrons injected from the compact nebula due to electron NRI along magnetic field lines.

\section{Acknowledgments}.

The authors thank Keping Qiu, Zhi-Yu Zhang, and Ruo-Yu Liu for helpful advice.
Y.C.\ acknowledges the support from NSFC under grants 12173018, 12121003, and 12393852.


\bibliography{sample631}{}

\bibliographystyle{aasjournal}

\end{document}